\documentstyle[emulapj]{article}

\font\tenbg=cmmib10 at 10pt
\def \rvecmu{{\hbox{\tenbg\char'026}}}

\begin{document}

\title{Three-dimensional Simulations of Disk Accretion to an
Inclined Dipole: II. Hot Spots and Variability}

\author{M.M.~Romanova}
\affil{Department of Astronomy, Cornell University, Ithaca, NY
14853-6801;~ romanova@astro.cornell.edu}

\author{G.V.~Ustyugova}
\affil{Keldysh Institute of Applied Mathematics,
   Russian Academy of Sciences, Moscow, Russia;~
ustyugg@spp.Keldysh.ru}

\author{A.V.~Koldoba}
\affil{Institute of Mathematical Modelling,
   Russian Academy of Sciences, Moscow, Russia;~koldoba@spp.Keldysh.ru}

\author{R.V.E.~Lovelace}
\affil{Department of Astronomy, Cornell University, Ithaca, NY
14853-6801; ~RVL1@cornell.edu }

\medskip


\begin{abstract}

    The physics of the ``hot spots" on stellar surfaces and the
associated variability of accreting magnetized rotating stars is
investigated for the first time using fully three-dimensional
magnetohydrodynamic simulations.
  The magnetic
moment of the star $\rvecmu$ is inclined relative to its rotation
axis $\bf{\Omega}$ by an angle $\Theta$ (we will call this angle
the ``misalignment angle") while the disk's rotation axis is
parallel to $\bf{\Omega}$. A sequence of misalignment angles was
investigated, between $\Theta=0^\circ$ and $90^\circ$.
     Typically at small $\Theta$ the
spots are observed to have the shape of a bow which is curved
around the magnetic axis, while at largest $\Theta$ the spots have
a shape of a bar, crossing the magnetic pole. The physical
parameters (density, velocity, temperature, matter and energy
fluxes, etc.) increase toward the central regions of the spots,
thus the size of the spots is different at different values of
these parameters.
       At relatively low density and temperature,
   the spots occupy approximately $10-20\%$
of the stellar surface, while at the
   highest values of these parameters
this area may be less than $1\%$  of the area of the star. The
size of the spots increases with the accretion rate.
      The  light curves were calculated
for different $\Theta$ and inclination angles of the disk $i$.
They show a range of variability patterns, including one
maximum-per-period curves (at most of angles $\Theta$ and $i$),
and two maximum-per-period curves (at large $\Theta$ and $i$). At
small $\Theta$, the funnel streams may rotate faster/slower than
the star, and this may lead to quasi-periodic variability of the
star. The results are of interest for understanding the
variability and quasi-variability of Classical T Tauri Stars,
millisecond pulsars and cataclysmic variables.

\end{abstract}

\section{Introduction}

      In accreting magnetized
stars the inflowing matter is channelled to regions near the
magnetic poles of the star, forming ``hot spots" on the star's
surface (e.g., Ghosh \& Lamb 1979; Uchida \& Shibata 1985;
Camenzind 1990; K\"onigl 1991; Shu et al. 1994; Hartmann et al.
1994).
     The hot  spots
form under conditions in which the star's magnetic field is strong
enough to form a magnetosphere.
    Examples include
  classical T Tauri stars
(e.g., Herbst et al. 1986; Bouvier \& Bertout 1989; Johns \& Basri
1995;  1998;  Petrov, et al. 2001a,b; Alencar \& Batalha 2002),
cataclysmic variables (e.g., Wickramasinghe, Wu, \& Ferrario 1991;
Livio \& Pringle 1992; Warner 1995; Warner 2000), X-ray pulsars
(e.g., Ghosh \& Lamb 1979; Tr\"umper et al. 1985; Bildsten et al.
1997) and millisecond pulsars (e.g., Chakrabarty, et al. 2003).
   These objects have different dimensions and
magnetic field strengths, but the underlying physics of the
magnetospheric accretion and hot spots' properties are expected to
be similar.

\begin{figure*}[t]
\epsscale{2.} \plotone{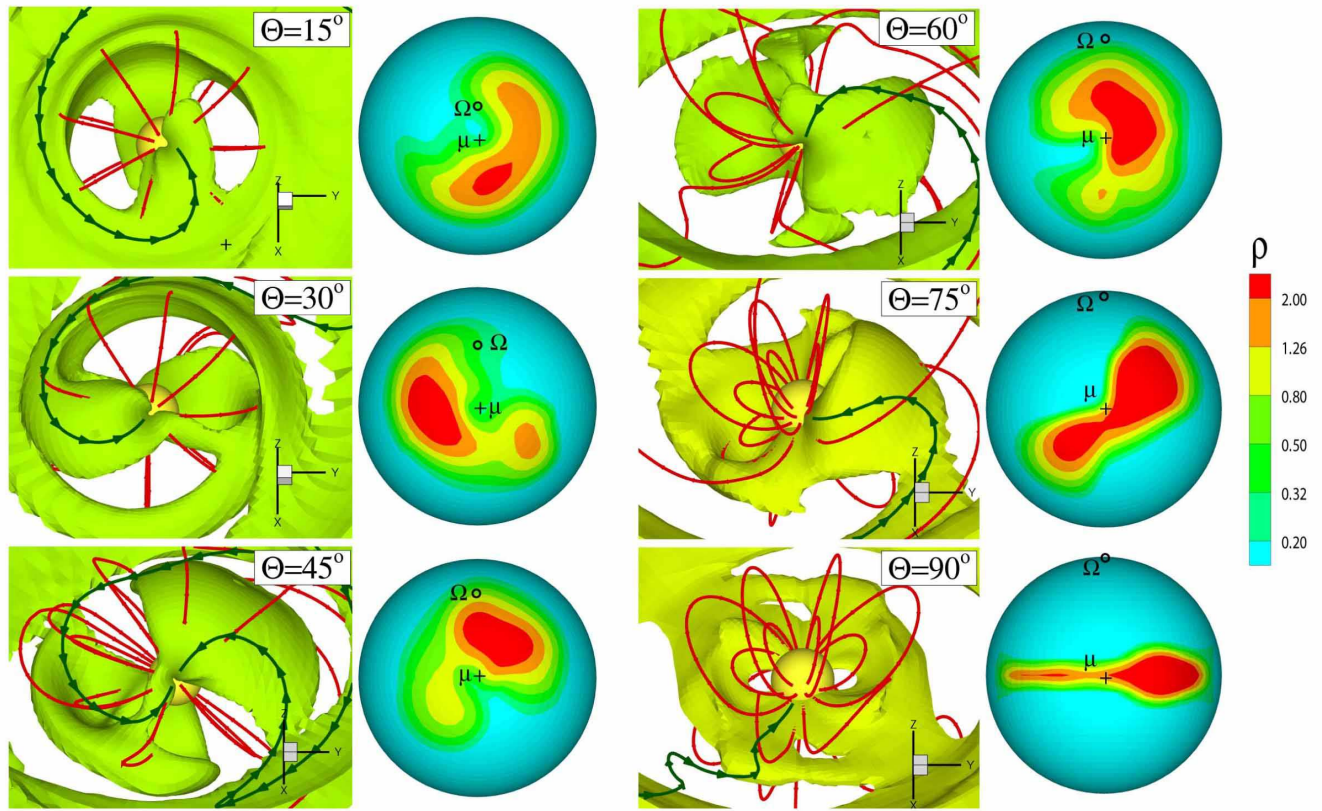} \caption{Matter flow near the star
and hot spots for the case of a relatively warm disk ($T_d=0.03$).
The square panels show the surfaces of the density in the
magnetospheric flows (green color corresponds to $\rho=0.3-0.4$,
while yellow-green color to $\rho=0.6$). Red lines are sample
magnetic field lines; the dark-green lines show the sample
streamlines of matter flow. The circular panels show density
distribution in the hot spots for different misalignment angles
$\Theta$.
 For
$\Theta=15^\circ, 30^\circ$ and $45^\circ$, the magnetic axis is
directed towards the observer, while for $\Theta=60^\circ,
75^\circ$ and $90^\circ$, it is directed at an angle $45^\circ$
relative to the observer. It is shown by the crosses in the
circular panels. The direction of the rotation axis is shown with
the open circles.} \label{Figure 1}
\end{figure*}

    The observed variability/quasi-variability of
these objects can be due to a number of possible phenomena in the
magnetosphere (see, e.g., Bouvier 2003; Petrov 2003).
 One of the  time-scales may be associated with
the hot spots.
    In spite of the large volume
of the observational data on variability of these stars,
relatively little is known about the properties of the hot spots.

For example, in classical T Tauri stars  (CTTS), hot spots
radiation is associated with veiling of the continuum radiation
and was analyzed for a number of CTTS.
    Herbst \& Koret (1988) and Bouvier et al.
(1986, 1993) estimated parameters of the hot spots assuming a
model with black body radiation, while Lamzin (1995), Calvet \&
Gullbring (1998), Gullbring et al. (2000), and Ardila \& Basri
(2000) used models of radiation from a shock wave. They concluded
that the area of the star covered with the hot spots in different
CTTS varies from $0.3\%$ to $20\%$ depending on the accretion rate
and other factors.
   In all models, however, the hot spots were considered to be
homogeneous, that is, of the same density and temperature.
     Further, some of
the models did not take into account the effect of limb-darkening.
   Additionally,
the shape and location of the hot spots were not known because
they can be derived only from a three-dimensional analysis.
Similar problems and questions appear during analysis of X-ray
pulsars and cataclysmic variables.
 Thus, it is important to study the properties of the hot spots
 in greater detail.
For analysis of the hot spots we use results of our previous
three-dimensional simulations (Romanova et al. 2003, hereafter -
R03), and also perform new simulations at a lower temperature in
the disk and at a variety of parameters of the star and the
accretion flow.

The main questions
which can be answered are: (1) What are the shapes of the hot
spots? (2) What is the distribution of physical parameters
(density, temperature) in the spots? (3) What is the area covered
by the spots at different physical parameters? (4) What are the
observed light curves at different misalignment angles $\Theta$
and different inclination angles of the disk $i$?

  In \S 2 of the paper we  describe
the underlying model and dimensional examples for CTTS and
millisecond pulsars.
    In \S 3 we discuss the expected  physical properties of
 the hot spots.
    In \S 4 we calculate the intensity of
radiation from the hot spots and the associated light curves.
    In \S 5 we discuss dependence of results on different
parameters, limitations of the model and future work.
    \S 6 gives the summary of our work.

\begin{figure*}[t]
\epsscale{2.} \plotone{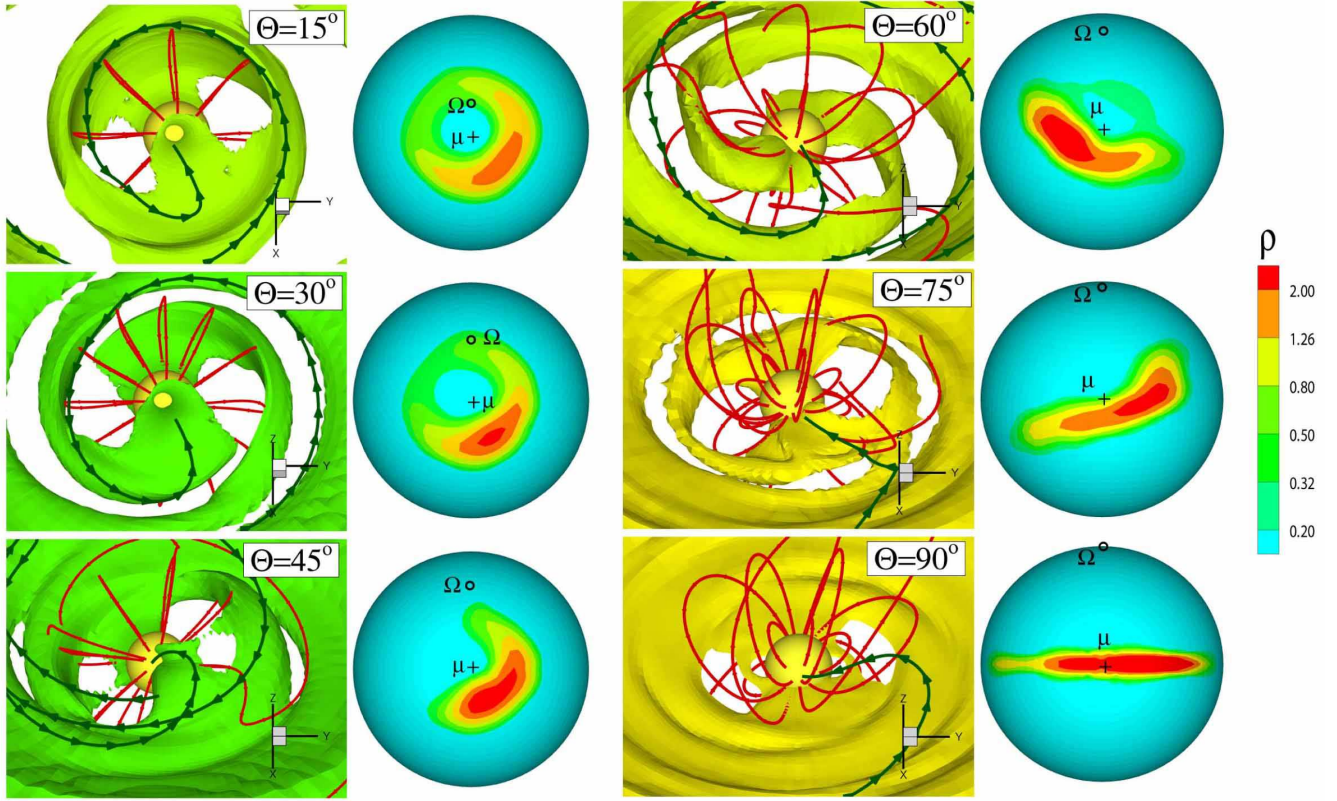} \caption{Same as Figure 1 but for
cooler disk ($T_d=0.01$).} \label{Figure 2}
\end{figure*}

\section{The Model and Reference Values}

  Below we briefly describe the numerical model of R03, initial and
boundary conditions, and  reference values for CTTS and
millisecond pulsars.

\subsection{The Model}

Disk accretion to a rotating star with an inclined dipole magnetic
field was investigated with three-dimensional MHD simulations.
   The magnetic moment $\rvecmu$ of the dipole is
inclined to the star's rotation axis ${\bf \Omega}$ by an angle
$\Theta$. The rotation axis of the star coincides with the
rotation axis of the disk.
   The value of the magnetic moment
$\mu=|\rvecmu|$ is restricted by computer resources.
   It is difficult to calculate in three dimensions
large magnetospheres, where the
  magnetospheric radius $R_m$ is much larger than the
star's radius $R_*$.
    This is because of the strong variation
($\propto 1/R^3$) of the dipole magnetic field.
    In R03, simulations were done for $R_m/R_*\approx 2-5$.
These values are appropriate for many  Classical T Tauri Stars
(CTTS) and millisecond pulsars. For  larger magnetospheres, the
inner boundary may be interpreted as an intermediate layer of
magnetosphere.

 A Godunov-type numerical code was used to solve the
full system of ideal MHD equations in three-dimensions space
written in a ``cubed sphere'' coordinate system rotating with the
star (Koldoba et al. 2002; R03).
     We use  a reference frame $(X,Y,Z)$ rotating with
the star. This frame is oriented such that the $Z$ axis is aligned
with the star's rotation axis and vector $\rvecmu$ is in the
$(X,Z)$ plane.

 The quasi-equilibrium {\it initial conditions} were proposed and
tested in axisymmetric model in Romanova et al. (2002 - hereafter
R02) and in three dimensions in R03. These initial conditions take
into account initial balance of gravitational, pressure gradient
and centrifugal forces. Although this initial condition does not
include the magnetic field, it avoids the possible initial
discontinuity of the poloidal magnetic field lines at the boundary
between the disk and corona (corona above the disk rotates with
the angular velocity of the disk). Test simulations with a
non-rotating corona show strong magnetic braking and accretion of
the disk with a speed close to the free-fall speed (see Hayashi,
Shibata \& Matsumoto 1996, some runs of Miller and Stone 1997).
The new initial conditions decrease this initial magnetic braking
dramatically, but not completely. There is a residual small
magnetic braking which determines a slow inward accretion of
matter to the star. This accretion is used as a source of
accreting matter in most of simulations. To check the validity of
this approach, we added simplified ``alpha" viscosity to the code,
where only the largest terms were taken into account (as in R02)
and performed a set of simulations at $\Theta=15^\circ$ and at a
variety of $\alpha$ parameters (see \S 5.1). Simulations have
shown that the velocity of the accretion flow induced by the
residual magnetic braking corresponds to viscous flow  with
$\alpha \approx 0.01-0.02$. These initial conditions allowed us to
observe for the first time the magnetospheric funnel flows from
the disk to the star in two and three dimensions  and to
investigate them in detail (see R02, and R03).

We consider the case when the star rotates relatively slowly with
angular velocity $\Omega_*\approx 0.04 \Omega_{K*}$, where
$\Omega_{K*}=\sqrt{GM/R_*^3}$ - is the brake-up  angular velocity
of the star.
    For the case of T Tauri stars,
this corresponds to a  slow rotation rate ($T\approx 9.4~ {\rm
days}$ for the parameters used in R02). This rotation velocity was
chosen simply because it was the main case of R03. Test cases with
faster rotation of the star were also performed.

We consider two main cases: (1) a relatively hot disk where the
flow to the star is subsonic, and (2) a cooler disk, where the
flow is supersonic.  The important parameters are the initial
densities in the disk $\rho_d$ and corona $\rho_c$, and the
initial temperatures in the disk $T_d$ and corona $T_c$.
 These values are determined at the fiducial point at the
boundary between the disk and corona (at the inner radius of the
disk) such that to support pressure balance in vertical direction.
In two-dimensional simulations (R02) we used such parameters that:
$\rho_c=0.01 \rho_d$, $T_d=0.01 T_c$ and obtained supersonic
funnel streams, because the sound speed is relatively low,
$c_s\approx 0.1 v_K$. However, in 3D simulations in R03 we were
able to cover only a ``softer" range of parameters: $\rho_c=0.03
\rho_d$, $T_d=0.03 T_c$, when the disk has higher temperature and
$c_s\approx 0.17 v_K$.
 In this paper we  use
 results of R03 and newer runs  with lower temperature in the disk
 ($c_s\approx 0.1 v_K$).

\begin{figure*}[t]
\epsscale{2.0} \plotone{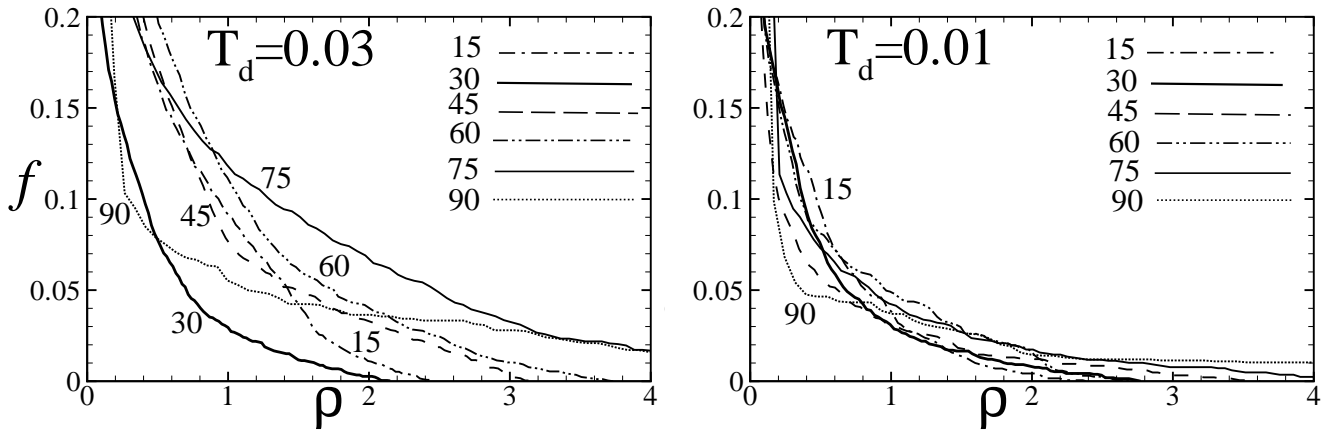} \caption{Fraction of the star's
surface area ${\it f}$ covered by the hot spots with density
larger than $\rho$ for different misalignment angles $\Theta$. The
left panel shows the case of warmer disk, $T_d=0.03$, while the
right panel the case of cooler disk, $T_d=0.01$.} \label{Figure 3}
\end{figure*}

The {\it boundary conditions} are similar to those in R03. At the
stellar surface there are  ``free" boundary conditions to the
density and pressure. The poloidal components of the magnetic
field are determined from the fact that the star is treated as a
perfect conductor rotating at the rate ${\bf \Omega}=\Omega
\hat{\bf z}$. There is however a ``free" condition to the
azimuthal component of the magnetic field: ${\partial(R
B_\phi)}/{\partial R}=0$ so that magnetic field lines have a
``freedom" to bend near the stellar surface.
    In the reference frame rotating with
the star the flow velocity  is adjusted such that to be parallel
to ${\bf B}$ at $R=R_*$. This inner boundary condition is valid
when the flow is subsonic. In the supersonic case there will be
standoff shock, but it requires separate investigation. From the
other side, these boundary conditions are valid for any of these
cases if the boundary is not the surface of the star, but
corresponds to some intermediate layers of the magnetosphere,
while the real surface of the star is at smaller radii. Thus, the
obtained parameters of the hot spots reflect the distribution of
these parameters in the cross-section of the funnel streams.

      At the outer boundary $R=R_{\rm max}$,
free boundary conditions are taken for all variables. We
investigated different types of boundary conditions for the
magnetic field, but did not observe a significant difference,
because the magnetic field of the dipole decreases very rapidly
with distance and the difference is much less dramatic compared to
monopole magnetic field (Ustyugova et al. 1999).
      We also investigated the possible influence of the outer
boundary conditions by running cases where the simulation regions
had $R_{\rm max}/R_*=14$, $40$, and $194$.
     We find the same results except for the case
$R_{\rm max}/R_*=14$ where the accretion rate decreases too fast
because the reservoir of matter in the disk is too small.
     Results for
the medium and large regions are very close, so that we took
$R_{\rm max}/R_*=40$ as a standard size of the simulation region.

\subsection{Dimensionless Variables and Reference Values}

    We use the
following dimensionless variables: the length scale $r'=r/R_0$,
the fluid velocity ${\bf v}'={\bf v}/v_0$, the density
$\rho'=\rho/\rho_0$, the magnetic field $B'=B/B_0$, the pressure
$p'=p/p_0$, the temperature $T'=T/T_0$ and time $t'=t/t_0$.
    The `zero' subscript variables are
dimensional reference values, which are different for different
objects.

   The reference values are determined as follows:
   the unit of distance $R_0$ is taken
to be an initial inner radius of the disk, $R_0=(R_d)_{t=0}$, so
that $r^{'}=1$ at this radius. The star has radius $R_*=0.35 R_0$.
    The reference velocity is the Keplerian
velocity at $R_0$, $v_0 =(GM/R_0)^{1/2}$, where $\omega_0 =
v_0/R_0$ is the angular velocity.
   The reference time is $t_0=R_0/v_0$.
   However, in discussing our results  we measure
time in units of the Keplerian period of the disk, $P_0=2\pi t_0$.
   The reference magnetic field $B_0$ is the initial
magnetic field strength at $r=R_0$.
    The reference density is taken to
be  $\rho_0 = B_0^2/v_0^2$.
    The reference pressure is $p_0=\rho_0 v_0^2$.
The reference temperature is $T_0=p_0/{\cal R} \rho=v_0^2/{\cal
R}$, where ${\cal R}$ is the gas constant.
    The reference accretion rate is $\dot M_0
= \rho_0 v_0 R_0^2$.
    The reference energy flux is $\dot
E_0=\rho_0 v_0^3 R_0^2$. The reference value for the black-body
temperature of the hot spots is $(T_{eff})_0=(\rho_0
v_0^3/\sigma)^{1/4}$, where $\sigma$ is the Stephan-Boltzmann
constant.

Subsequently, we drop the primes on the dimensionless variables
and show dimensionless values in most of the figures. Two types of
initial conditions for the disk in dimensionless variables are
considered: $\rho_d=1$, $\rho_c=0.03$, $T_d=0.03$ and $T_c=1$ for
the warmer  disk, and $\rho_d=1$, $\rho_c=0.01$, $T_d=0.01$ and
$T_c=1$ for the cooler disk. Below we discuss dimensional examples
for stars with relatively small magnetospheres, CTTS and
millisecond pulsars.

\begin{figure*}[t]
\epsscale{2.0} \plotone{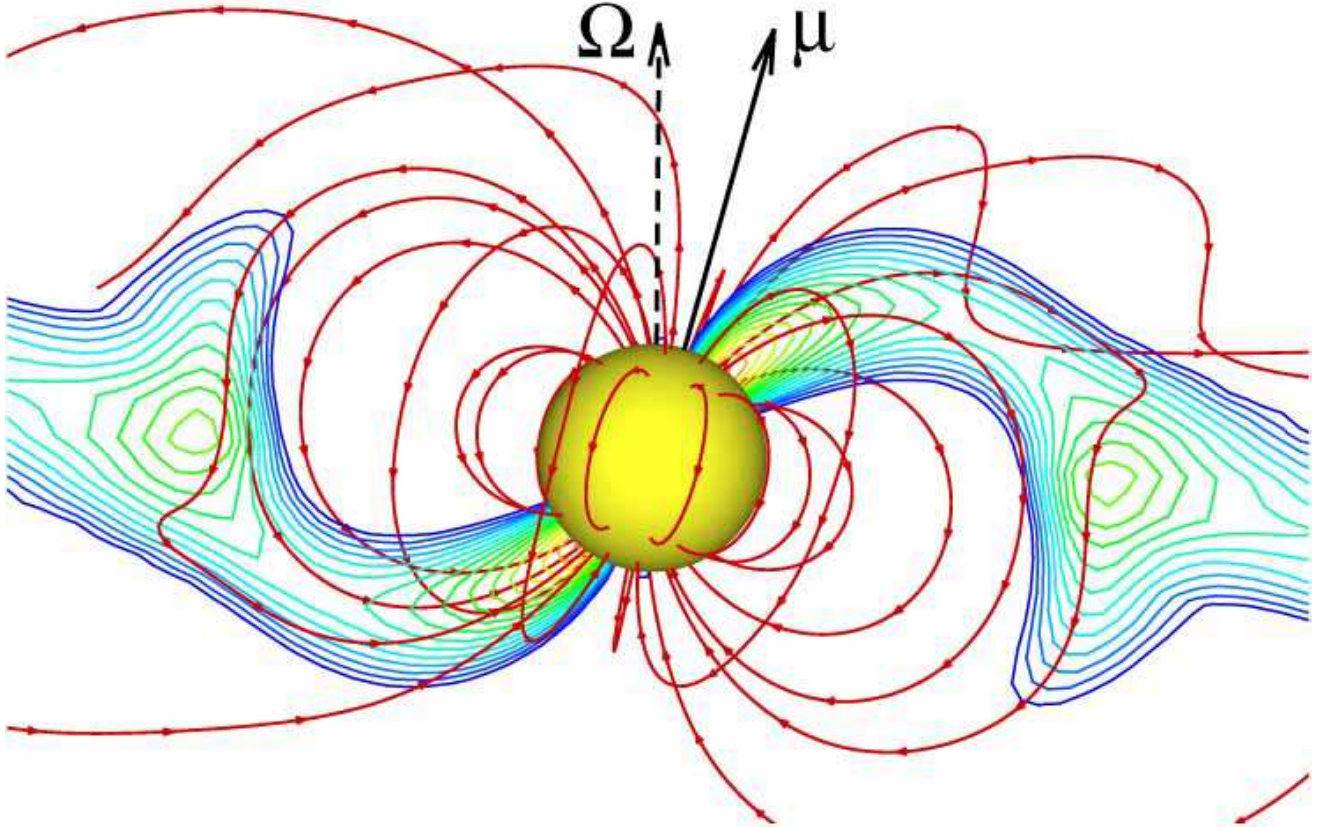}\caption{An $X-Z$ slice through the
middle of the funnel stream at $\Theta=15^\circ$.  The contour
lines show the plane cross-section of the density distribution
inside the funnel stream. The density changes exponentially from
$\rho=0.2$ (blue) to $\rho=2.0$ (red). The density in the corona
above the disk is $\rho=0.01-0.02$.
 Red
lines with arrows show selected magnetic field lines. The magnetic
moment $\rvecmu$, and the rotation axis $\bf\Omega$ are shown.}
\label{Figure 4}
\end{figure*}

\subsection{Reference Values for Classical T Tauri Stars}

Here, we discuss the numerical parameters for a typical CTTS.
   We take the mass  of the star
to be $M_*=0.8 M_\odot$, and its radius $R_*=1.8 R_\odot$.
    The magnetic field at the
surface of the star is assumed to be $B_*=10^3~{\rm G}$.
    The reference value of length  is  $R_0=(1/0.35) R_* \approx 2.86 R_*
\approx  3.6\times 10^{11}~{\rm cm}$.
    The size of the simulation region
corresponds to $R_{\rm max}\approx 40 R_*\approx 0.34~{\rm AU}$.
   The reference  velocity  is
$v_0 \approx 1.9\times 10^7~{\rm cm/s}$.
   The period of Keplerian
rotation of the inner radius of the disk is $P_0 \approx 1.38~
{\rm days}$.
    We consider that the star rotates relatively slowly,
$P_*=9.4~{\rm days}$ which corresponds to many CTTSs.
   Simulations for
higher rotational velocities did not bring principally new results
(excluding the propeller stage of course, which we do not discuss
here).
    The reference magnetic field is  $B_0=B_*(R_*/R_0)^3
\approx 42.7 ~{\rm G}$.
            The reference density is
$\rho_0=4.9\times 10^{-12}~{\rm g/cm^3}$ or $n = 3.06\times
10^{12}~{\rm cm^{-3}}$ which is typical for T Tauri star disks
(see, e.g., Hartmann et al. 1998).
    The reference temperature is
$T_0=4.5\times 10^6~{\rm K}$ which corresponds to typical
temperature in the corona and temperature $T_d=4.5\times10^4~{\rm
K}$ in the innermost part of the disk for the run typical for CTTS
($T_d=0.01$, $T_c=1$). The reference mass accretion rate is $\dot
M_0 \approx 1.2\times 10^{19}~{{\rm g/s}} \approx 1.9\times
10^{-7}~ {\rm M_\odot}/{\rm year}$.
   The reference value
for the energy flux is $\dot E_0\approx 4.5\times 10^{33}~{\rm
ergs/s}$ and for black-body temperature: $(T_{eff})_0=5.0\times
10^3~{\rm K}$.
     The dimensionless values of the
intensity  in the Figures 7-9 are $J\approx 2\times 10^{-3} -
1.7\times 10^{-2}$ which corresponds to the dimensional values $J
\dot E_0\approx 9.0\times 10^{30} - 7.6\times 10^{31}~{\rm
ergs/s}$.

\subsection{Reference Values for Millisecond Pulsars}

We take the mass of the neutron star to be $M_*=1.4~{\rm M_\odot}=
2.8\times 10^{33}~{\rm g}$,  its radius $R_*=10~{\rm km}=10^6~{\rm
cm}$, and the surface magnetic field $B = 10^8~{\rm G}$.
   The reference length scale  is $R_0\approx 2.86 R_*=2.86\times
10^6~{\rm cm}$. The reference velocity is $v_0=8.1\times 10^9~{\rm
cm/s}$. The reference period is $P_0=2.2\times 10^{-3}~{\rm s}$.
    In application to millisecond pulsars often the
frequency is analyzed.
    The reference frequency is $\nu_0=1/P_0\approx 454~{\rm
Hz}$.
   The reference magnetic field is  $B_0=B_*(R_*/R_0)^3 \approx
4.3\times 10^6~{\rm G}$.
            The reference density is
$\rho_0=2.8\times 10^{-7}~{\rm g/cm^3}$.
    The reference temperature
is $T_0=7.8\times 10^{11}~{\rm K}$.
   The reference mass accretion
rate is $\dot M_0 \approx 1.85\times 10^{16}~{{\rm g/s}} \approx
2.9\times 10^{-10}~{\rm M_\odot}/{\rm year}$.
    The reference value
for the energy flux is $\dot E_0\approx 1.2\times 10^{36}~{\rm
erg/s}$. The reference black-body temperature is
$(T_{eff})_0=4.0\times 10^6~{\rm K}$.
   The dimensionless values of the intensity  in Figures 7-9
are $J\approx 2\times 10^{-3} - 1.7\times 10^{-2}$ which
corresponds to dimensional values $J \dot E_0\approx 2.4\times
10^{33} - 2.0\times 10^{34}~{\rm ergs/s}$.

\section{Physical Properties of the Hot Spots}

In this section we discuss the connection between the shape of the
funnel streams with that of the hot spots (\S 3.1), properties of
matter along the funnel streams (\S 3.2), and  properties of the
hot spots (\S 3.3)

\subsection{Funnel Streams Versus Hot Spots}

The shape and other properties of the hot spots are determined by
those in the cross-section of the incoming funnel stream.  In R03
we observed that the shapes of the funnel streams are different
for different misalignment angles $\Theta$ and may be complicated.
At small misalignment angles, $\Theta\lesssim 30^\circ$, matter
typically accretes in two streams.
      For ``medium" angles,
$30^\circ\lesssim \Theta\lesssim 60^\circ$, the streams are often
split into several streams. At even larger angles, $\Theta\gtrsim
60^\circ$, matter again accretes in two streams, but they have
different shapes compared to those at small $\Theta$. The streams
tend to settle in the particular location, close to the
$\mu-\Omega$ plane, after few rotation periods $P_0$. They settle
faster for cooler disks and large $\Theta$. We chose one moment of
time $t=5 P_0$, at which many streams settled, and investigate
spots in detail for this time.

 Figure 1 shows the funnel streams and corresponding hot spots
at the surface of the star at time $t=5 P_0$. Only a small part of
the simulation region is shown in order to resolve the inner
regions of the funnel streams in greater detail. To show the shape
of the funnel streams, we have chosen for each plot a fixed
density level which is slightly different for different $\Theta$
and varies between $\rho=0.35-0.45$ for smaller $\Theta$ (green
color), and $\rho=0.6-0.7$ (yellow-green color) for larger
$\Theta$. At these density levels the size of the spots is
relatively large and they occupy approximately $20\%$ of the
star's surface area. One can see that at all $\Theta$ the density
of the spots increases towards their central regions. These
results correspond to simulations of warmer disk, $T_d=0.03$.

 Figure 2 shows similar results, but for the cooler disk,
 $T_d=0.01$.
One can see that the funnel streams and the spots are somewhat
similar. However, in this case (which is taken also for the
 time $t=5 P_0$) the streams and spots are located much closer
to their ``final", quasi-equilibrium position (which is downstream
of the $\mu-\Omega$ plane for slowly rotating star, - R03),
compared to the warmer disk. In both cases, of warmer and cooler
disks, the spots continue to wander around this quasi-equilibrium
position. However, in the case of the cooler disk the amplitude of
these oscillations is much smaller.

We calculated the filling factor of the spots ${\it f}$ (the
fraction of the star covered by the spots) for both cases at
different density levels $\rho$ and different $\Theta$. Figure 3
shows that at given $\Theta$ the value ${\it f}$ decreases with
$\rho$, that is, at larger densities the area covered by the spots
is smaller. In case of warmer disk (left panel), the size of the
spots strongly depends on $\Theta$, specifically at larger
densities, while in case of cooler disks this dependence is much
less prominent. At the density levels corresponding to the funnel
streams, the size of the spots is relatively large and they occupy
approximately $10-20\%$ of the star's surface area. The spots are
typically smaller for the cooler disk.

\subsection{Parameters Along the Funnel Stream}

To investigate parameters along the funnel stream, we consider the
case, $\Theta=15^\circ$, where the pattern of the streams is
relatively simple. Figure 4 shows the slice through the middle of
the funnel stream for $T_d=0.03$. One can see that matter
accumulates near the magnetosphere forming a relatively dense region
- a ring. When matter goes to the funnel stream, the density along
the flow first decreases,  but later increases due to the
converging of the magnetic field lines.  Figure 4 shows that the
density is largest in the interior regions of the stream and
decreases outward to much lower values.

 Figure 5 shows velocities
along $v_{||}$ and across $v_t$ the stream and Mach number ${\it
M}=v/c_s$, where $v$ is the total velocity. The left panel
corresponds to $T_d=0.03$. One can see that in the beginning of
the stream, the velocity across the stream $v_t$ is comparable
with $v_{||}$,  but it decreases while approaching the star.
  The  velocity does not reach
supersonic values, because the sound speed is relatively high and
also because the pressure gradient decelerates the flow.
  In case of cooler disk (right panel)
 the  $v_{||}$ component is larger, the Mach number is larger
than unity and flow is {\it supersonic}. Below we investigate the
properties of the hot spots for these two cases.

\subsection{Density, Velocity and Matter Flux in the Spots}

Figure 6 shows an example of distribution of different parameters
in the hot spots for  $\Theta=30^\circ$ for cases of subsonic flow
(top two rows) and supersonic flow (bottom two rows). One can see
that both, density $\rho$ and total velocity $v$  increase towards
the central regions of the spot. The distribution of the Mach
number ${\it M}=v/c_s$  is similar to  that of $v$. The Mach
number reaches values ${\it M}\gtrsim 2$ in the middle of the spot
in case of cooler disk.

The matter flux at each point ${\bf R}$ of the star's surface is
$$ F_M({\bf R}) = \rho ~\hat{\bf n}\cdot {\bf
v}~, \eqno(1)$$
 where $\hat{\bf n}=-\hat{\bf r}$ is the inward pointing
normal vector to the star's surface.
    The total mass accretion rate is
$$ \dot{M} = R_*^2\int d\Omega~F_M({\bf R})~, $$
 where $d\Omega$
is the solid angle element. Figure 6 shows that the $F_M$
distribution is similar to those of density and velocity. Note,
that in both, subsonic and supersonic cases the maximum values of
$F_M$ are approximately the same, which reflects the fact that the
accretion rate from the disk is approximately the same in both
cases.

\begin{figure*}[t]
\epsscale{1.8} \plotone{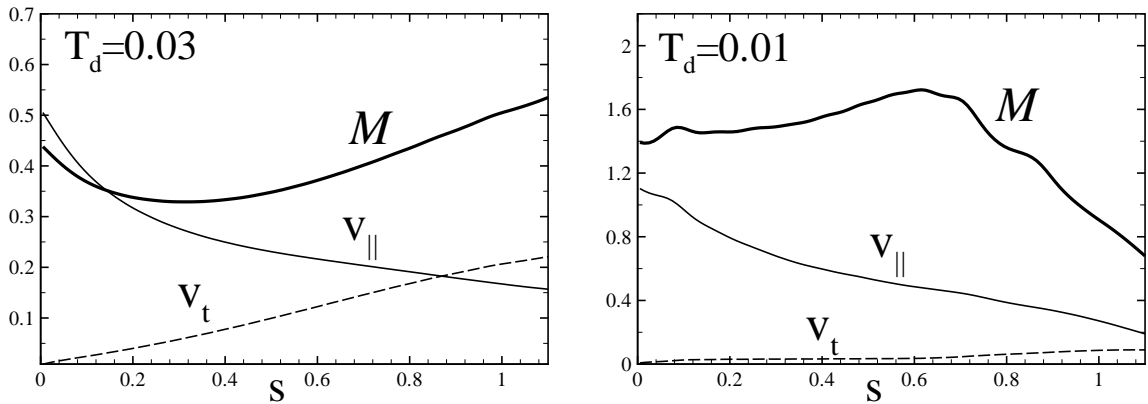}\caption{Distribution of velocities
along a magnetic field line going approximately through the middle
of the funnel stream obtained at $\Theta=15^\circ$. Variable $s$
is the linear distance along this magnetic field line, with $s=0$
at the surface of the star. Here, $v_{||}$ is the velocity along
the field line, $v_t$ is the velocity in the perpendicular
direction, ${\it M}=v/c_s$ is the Mach number based on the total
velocity of the flow. The left and right panels  show the examples
of the subsonic and supersonic flows correspondingly. }
\label{Figure 5}
\end{figure*}

\begin{figure*}[t]
\epsscale{1.6} \plotone{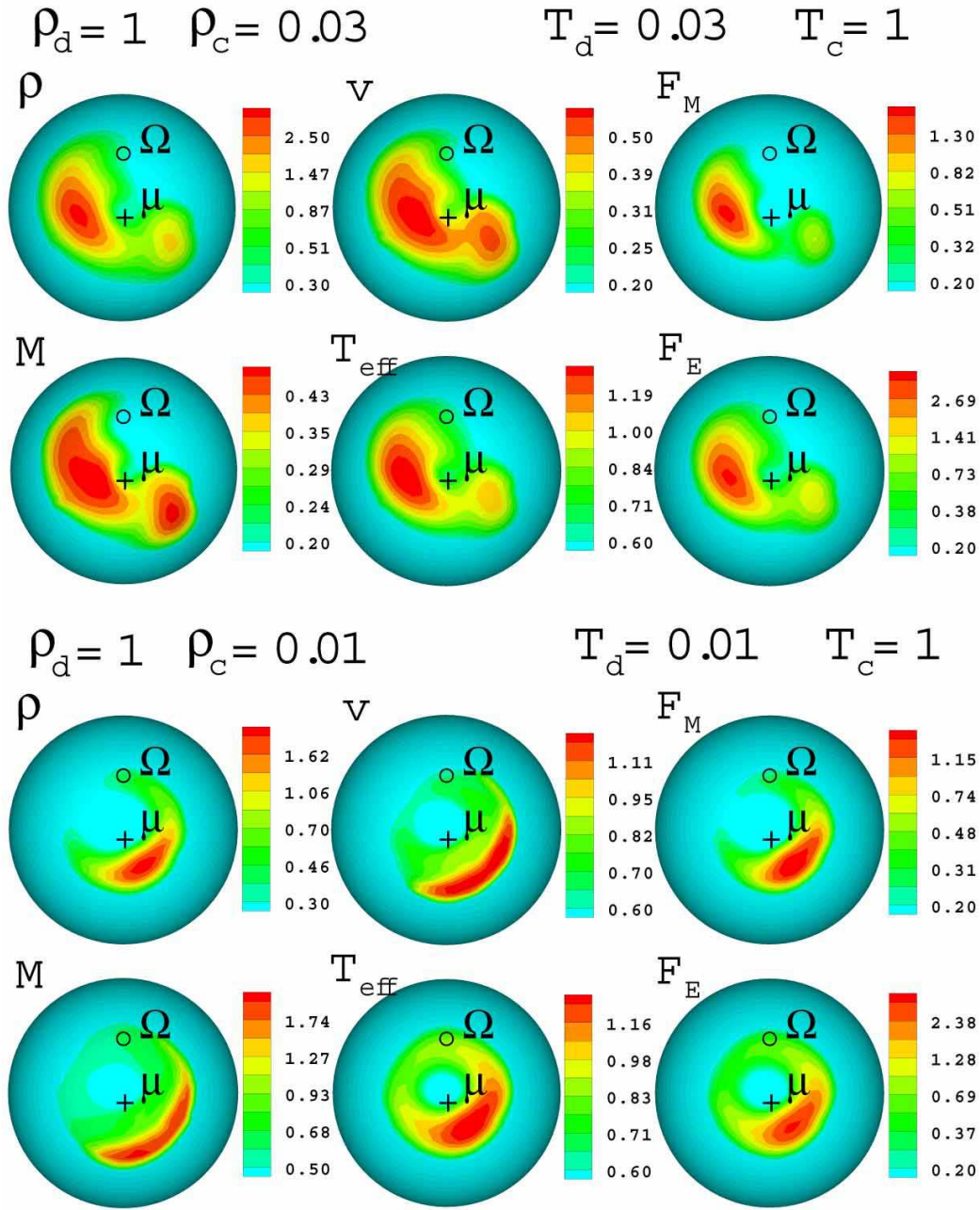} \caption{Distribution of different
physical parameters in the hot spots for a misalignment angle
$\Theta=30^\circ$ and for two types of initial conditions in the
disk and corona. $\rho$ is the density, $v$ is the total velocity,
$T_{eff}$ is the effective temperature, $\it M$ is the Mach number
of the flow, ${\it F}_M$ and ${\it F}_E$ are the fluxes of mass
and energy. The top two rows show the case of higher temperature
in the disk and subsonic flow, while the lower two rows show the
case of lower temperature in the disk and supersonic flow. }
\label{Figure 6}
\end{figure*}

\begin{figure*}[t]
\epsscale{1.8} \plotone{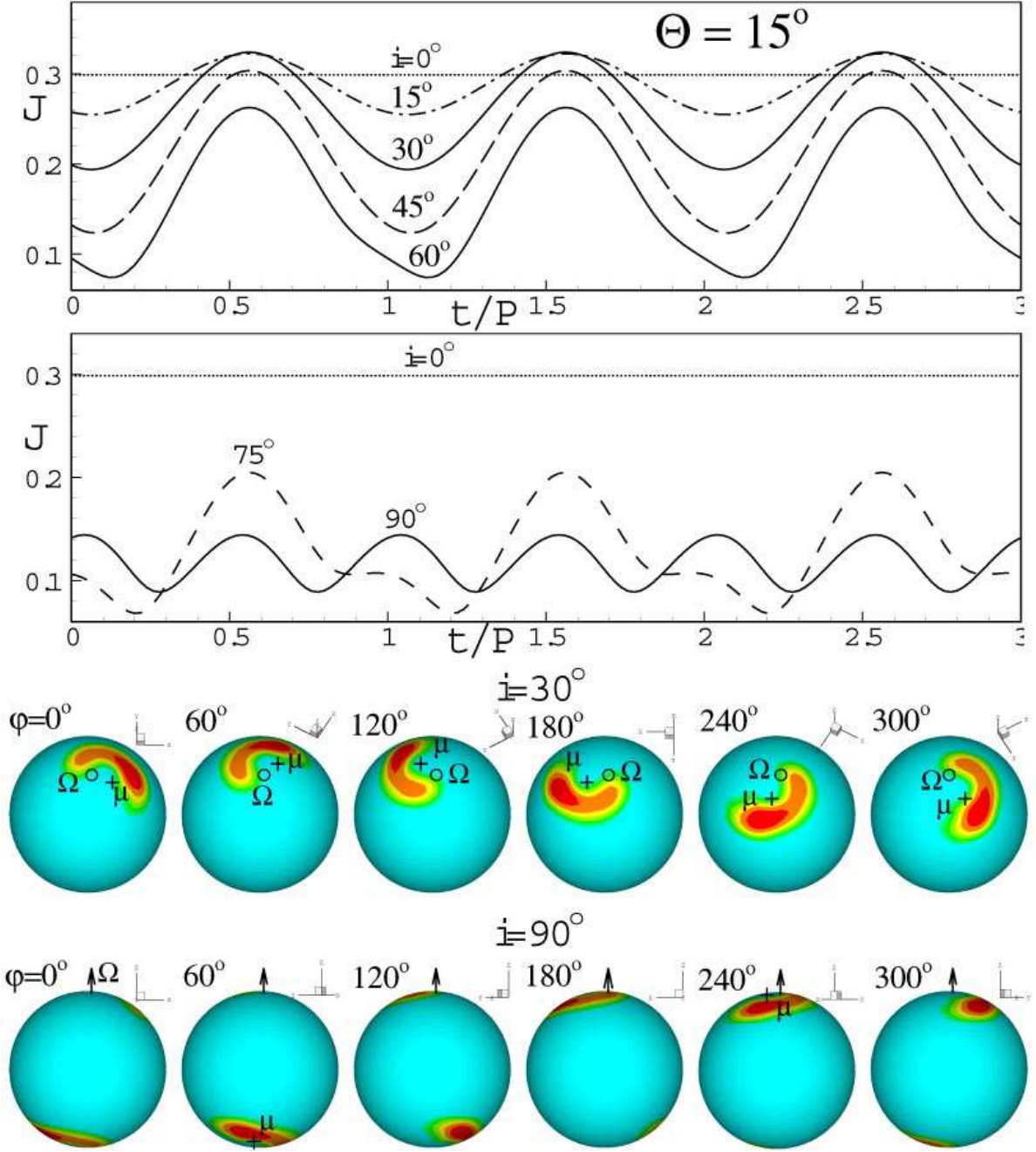} \caption{Light curves calculated
for small misalignment angle  $\Theta=15^\circ$ and for different
inclination angles $i$. Time $t$ is measured in periods of
rotation of the star $P$. The two bottom pictures show the
distribution of energy flux $F(R)$ in the hot spots at different
phases $\varphi$ of the star seen by observer at inclination
angles $i=30^\circ$ and $i=90^\circ$. Phases are shown for one
period of rotation.   The magnetic and rotation axes are marked
with a circle and a cross correspondingly.} \label{Figure 7}
\end{figure*}

\begin{figure*}[t]
\epsscale{1.8} \plotone{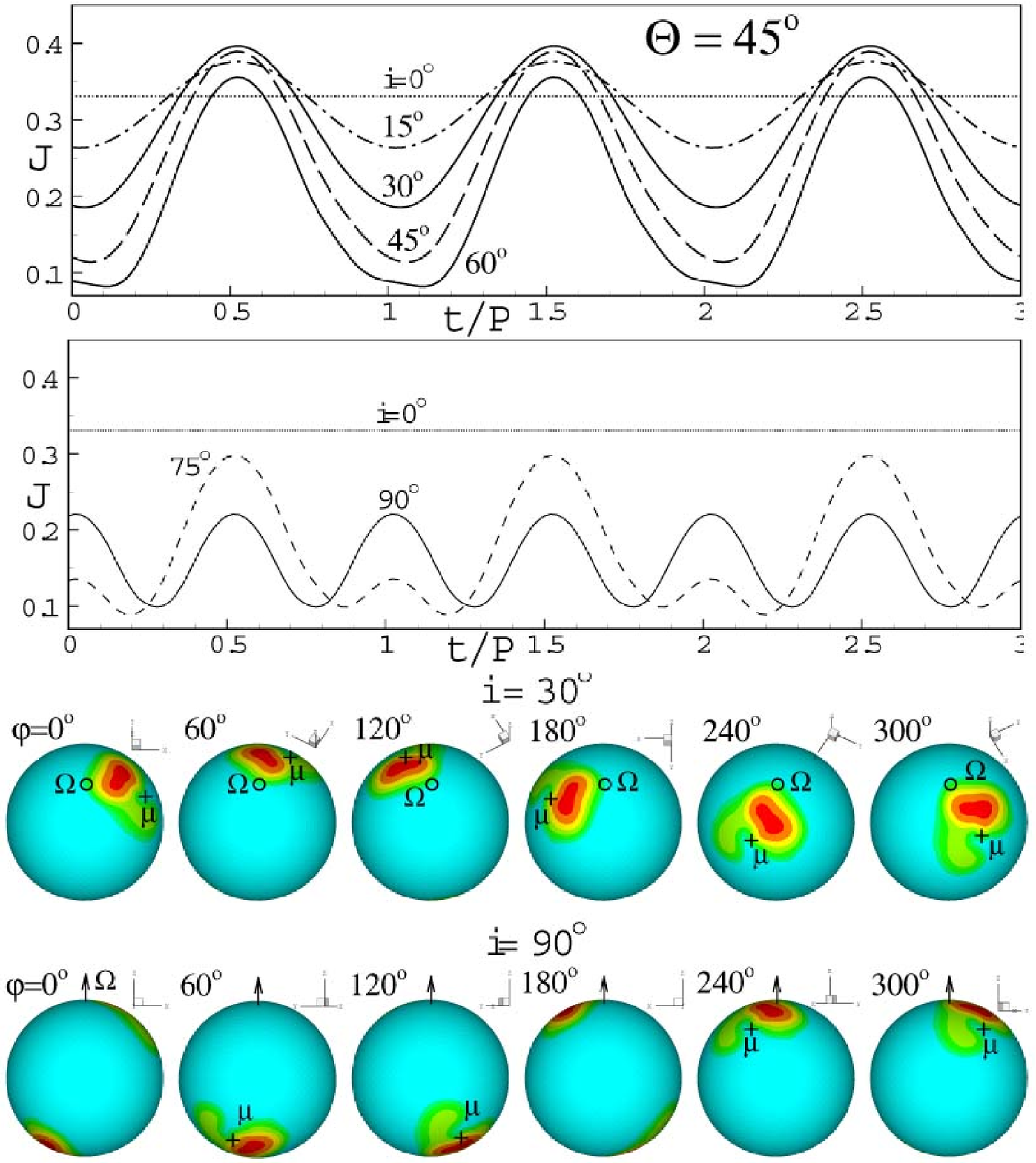} \caption{Light curves calculated
for medium misalignment angle  $\Theta=45^\circ$ and for different
inclination angles $i$. All features of the plot are the same as
in Figure 7.} \label{Figure 8}
\end{figure*}

\begin{figure*}[t]
\epsscale{1.8} \plotone{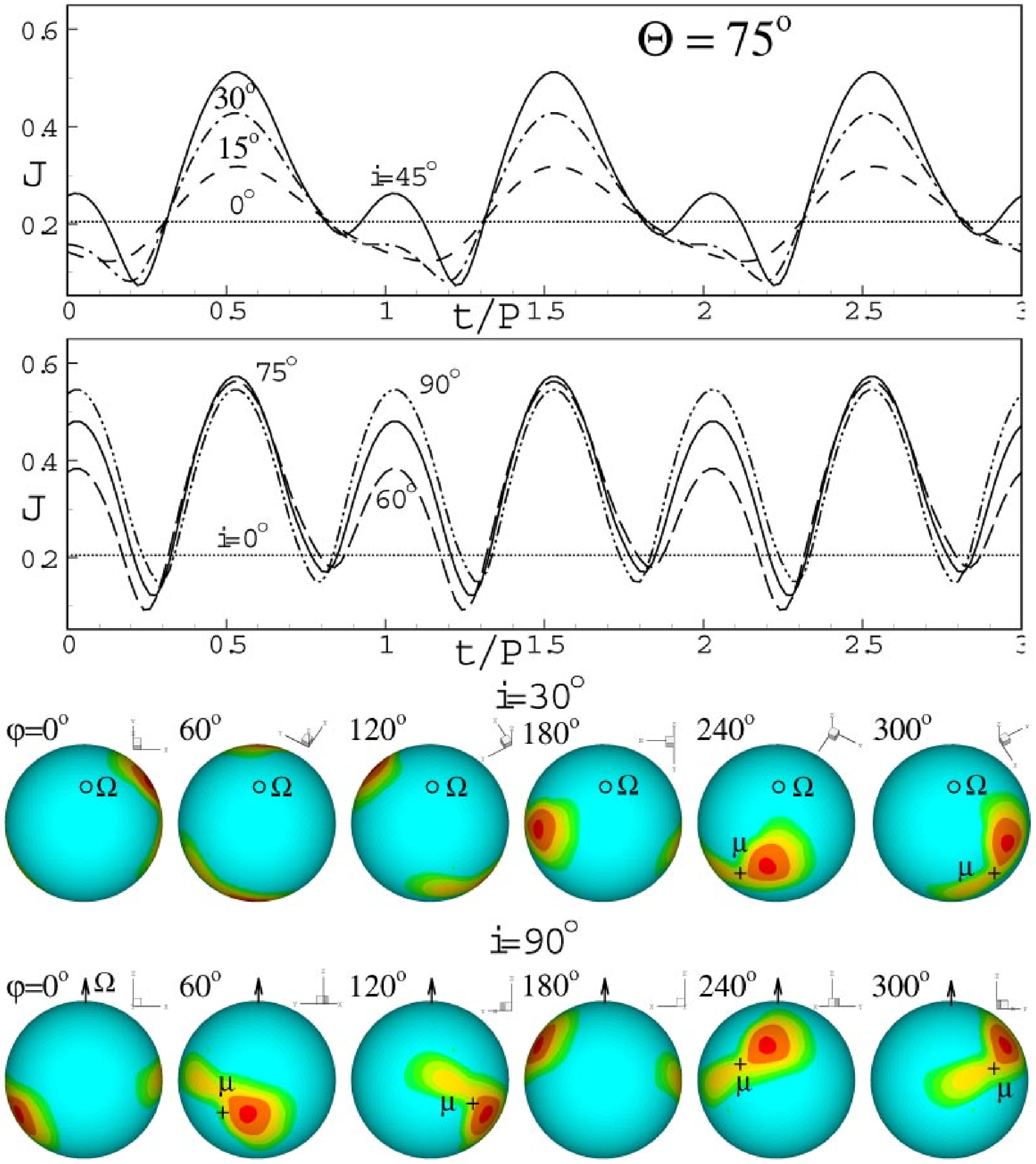} \caption{Light curves calculated
for a large misalignment angle  $\Theta=75^\circ$ and for
different inclination angles $i$. All features of the plot are the
same as in Figure 7.} \label{Figure 9}
\end{figure*}
\begin{figure*}[t]
\epsscale{1.2} \plotone{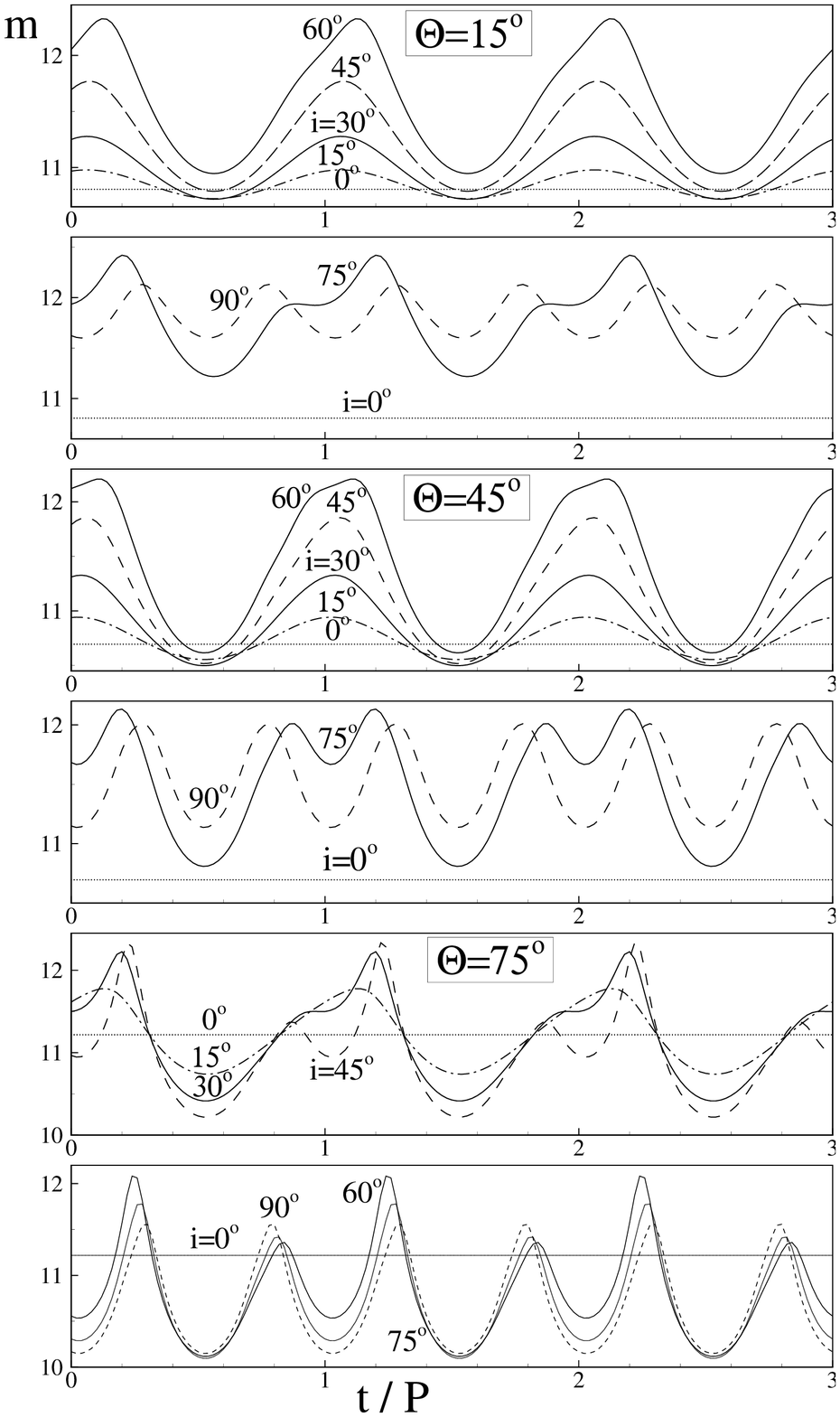} \caption{Light curves similar to
those of Figures 7-9, but the intensity is measured in stellar
magnitudes $m$. The top two panels correspond to
$\Theta=15^\circ$, the middle panels correspond to
$\Theta=45^\circ$ and the bottom two panels correspond to
$\Theta=75^\circ$.} \label{Figure 10}
\end{figure*}

\subsection{Energy Flux and Temperature in the Spots}

Matter flowing to the star in an accretion stream carries both
kinetic and thermal energies.
   The associated energy flux through the surface of the
star at the point ${\bf R}$ is:
$$ F_E({\bf R}) =  \rho~\hat{\bf
n}\cdot
  {\bf v}\left({1\over 2}{{\bf v}^2+
w}\right), \eqno(2) $$ where, ${\bf v}$ is the  velocity of plasma
relative to the surface of the star, $w=\gamma
(p/\rho)/(\gamma-1)$ is the specific enthalpy of the plasma, and
$\hat{\bf n} =-\hat{\bf r}$ is the inward pointing normal to the
surface of the star.

   The detailed physics of the hot
spots is complicated. To calculate the spectrum of radiation from
the hot spots an analysis of the radiation transfer is needed
(see, e.g., Lamzin 1998; Muzerolle, Hartmann, \& Calvet 1998;
Muzerolle, Calvet,  \& Hartmann 2001).
     This paper does not include an
analysis of the radiation transfer.
     We  obtain  approximate
temperature distributions in the hot spots based on overall energy
conservation.
    Namely, we assume that
energy  released in  the hot spots is due to the  energy flux of
equation (2).
   At the point ${\bf R}$ of the
surface of the star, this energy flux is considered to radiate as
a black body.
   Thus, $F_E({\bf R})=\sigma
T_{\rm eff}^4$, where $\sigma=5.67\times 10^{-5}({\rm erg/s})/{\rm
cm}^{2}/{\rm K}^{4}$ is a Stephan-Boltzmann constant and $T_{\rm
eff}$ is the effective black-body temperature of the radiation.
   Thus, we get
$$
     T_{\rm eff}({\bf R}) = \left[ {\rho~\hat{\bf n}\cdot
  {\bf v}\over \sigma}\left({1\over 2}{{\bf v}^2+
w}\right)\right]^{1/4}. \eqno(3) $$

  Figure 6 shows the distribution of   $F_{\rm E}$ and
  $T_{\rm eff}$  in the spots. Their shapes are similar to
  those of the density.
The distribution  of $T_{\rm eff}$ is important for understanding
the variability in different spectral bands. The spots are
expected to be smaller at higher $T_{\rm eff}$ and larger at lower
$T_{\rm eff}$.
 An indication of such distribution was possibly
observed in the CTTS object  BP Tauri, where  the area of the hot
spots was estimated to be different when different methods were
used.
    Errico, Lamzin \&
Vittone (2001) estimated the area to be $20\%$ of the surface of
the star.
    They used the spectral lines which are thought to originate
in the external regions of the funnel flow.
   Ardila \& Basri (2000)
estimated the area to be much smaller, $0.3\%$. However, they
modelled the UV continuum, which originates in the hottest region
of the spots. Note that in present analysis, we do not take into
account the radiation coming from the interior of the star.

\section{The Intensity of Radiation and Light Curves}

For calculation of the radiation intensity from the hot spots, we
suppose that all of the kinetic energy flux $F_E({\bf R}$ goes
into radiation.
 The energy, radiated by a unit area of the spot is
$$
   F_E({\bf R})= \int\limits_{\cos{\theta} > 0}
d\Omega ~f( {\bf R},{\bf m})~,
  \eqno(4)
$$
  where $f({\bf R},{\bf m})$ is the intensity of the radiation
from a unit area into the solid angle element $d\Omega$ in the
direction ${\bf m}$ with ${\bf m}\cdot \hat{\bf r} = \cos\theta$.
   The condition $\cos\theta>0$ corresponds to
the upper half space above the star's surface. For specificity we
consider $ f({\bf R},{\bf m}) = A({\bf R}) \cos{\theta}.$
     Thus we obtain the luminosity of unit area at
point ${\bf R}$, $$
       F_E({\bf R}) = 2\pi A({\bf R}) \int\limits_0^{\pi/2}
\sin{\theta}
        \cos{\theta}d{\theta}
  = \pi A({\bf R})~,
       $$
  and
$$ f({\bf R},{\bf m}) = {1\over \pi} {F_E({\bf R})\cos{\theta}}~.
\eqno(5) $$

     Consider now the intensity of radiation seen
by a distant observer.
    The unit vector from the star to the observer
is denoted  $\hat{\bf k}$.
    This vector makes an angle $\beta$ with the normal
to the star's surface, that is, $\cos \beta = \hat{\bf k}\cdot
\hat{\bf r}$.
     The  intensity of
radiation in the direction $\hat{\bf k}$ is obtained by
integrating over the stellar surface,
$$ J=\int {dS~ f({\bf
R},{\hat{\bf k}})} =\frac{1}{\pi} \int\limits_{\cos{\beta} > 0}
dS~ F_E({\bf R}) \cos{\beta}~, \eqno(6) $$
 where $dS$ is an element of the surface
area of the star.
  The condition $\cos{\beta} > 0$ or
$\beta < \pi/2$ corresponds to only the near side of the star
being visible.
       The rotation of a star  with
hot spots  leads to variations in the observed emission $J$.

\subsection{Variability Curves at Fixed position of Spots}

Hot spots constantly change their shape and position. However, in
many situations the changes are small (e.g., in case of the cooler
disk), so that  as a {\it first step} we suppose that the spots
have fixed location at the surface of the star and variability is
connected with rotation of the star relative to the fixed
observer.
  This approach helps to understand
different features of the light curves, associated with structure
and location of the hot spots at different $\Theta$. Examples of
actual variability curves are shown in \S 4.2.

 We consider one of the initial conditions (warmer disk, $T_d=0.03$),
 take hot spots at a single  moment of time, $t= 5
P_0$ (as in Figure 1),
    and  calculate from equation (6) the observed intensity
     $J(t)$ as the star rotates.
   Such  light curves were
calculated for  misalignment angles $\Theta = 15^\circ,~
45^\circ$, and $\Theta=75^\circ$, and for different inclination
angles of the disk $i$ with respect to the line of sight.
    The inclination angle of the disk is the angle between
the direction of ${\bf \Omega}$ and the direction to the observer
$\hat{\bf k}$. That is, $\cos i = \hat{\bf \Omega}\cdot \hat{\bf
r}~$. The results are shown at Figures 7 - 9.

Figure 7 shows results for a relatively small misalignment  angle
$\Theta=15^\circ$.
    The top panel shows the light curves  for
different inclination angles $i$.
    The bottom panel shows the orientation
of the star during one rotation period.
     The phases are shown for
two  cases: at $i=30^\circ$, which is a probable inclination
angle, and at $i=90^\circ$ which is edge on.

    Of course for $i=0^\circ$
there is no variability because the observer sees only one hot
spot which rotates  around the $\Omega$ axis.
     For larger inclination
angles, $i=15^\circ$, $i=30^\circ$, $i=45^\circ$,  dips appear in
the light curves which increase with increasing  of $i$.
   These
dips are connected with the fact that the hot spots come closer to
the edge of the star and are partially obscured.
    Under these conditions the star
will be observed as a variable star with one maximum per period.
    For $i= 60^\circ$,
the light curve becomes asymmetric and this asymmetry increases
for larger inclination angles $i$.
    This is connected with
the fact that the second spot, which was invisible at smaller $i$,
appears and starts to contribute to the luminosity.
    Thus, two intensity maxima
per stellar period are observed. The two maxima become of equal
amplitude at $i=90^\circ$.

Figure 8 shows similar plots, but for the misalignment angle
$\Theta=45^\circ$.
   The light curves
in this case are similar to those at smaller $\Theta$.
    However,  the shorter time-scale variability
appears at large inclination angles at $i \gtrsim 75^\circ$.

    Figure 9 shows the case of
a high misalignment angle, $\Theta=75^\circ$.
    For this and larger
values of $\Theta$, the double maxima (half-period) variability
appears for  many inclination angles starting from $i \gtrsim
30^\circ$.

The figures 7-9 show intensity in dimensionless units. This value
in dimensional units will be $J \dot E_0$, where $\dot E_0$ is the
dimensional energy flux (see \S 2.3 and \S 2.4). The value $J \dot
E_0$ gives the energy flux from the spots radiated to the
direction of the observer. This value will determine the observed
flux, when the distance to the object will be taken into account.
Thus, the X-ray flux observed e.g. from millisecond pulsars will
be proportional to $J$ and the expected variability curves will be
similar to those at the Figures 7-9. However, for CTTS it is
common to show the light curves on a logarithmic scale, using the
standard stellar magnitude values.  As long as this analysis can
be applicable to stars at different distances, we accept the
random null-point for calculation of $m$ which will be different
for different CTTS. We suppose that intensity $J=0.1$ corresponds
to $m=12$, and calculate the light curve using the standard
formulae: $m=12 - 2.512 (1.+\rm{log} J)$. Figure 10 shows
light-curves in stellar magnitudes for different $\Theta$ and $i$.
Note that the curves have different shapes which may be used for
estimating values of $\Theta$ and $i$.

\subsection{Variability Curves for Changing Spots}

Here, we present samples of actual variability curves. We  take
the data for all moments of time from the numerical simulations.
The star rotates slowly, with $\Omega_*=0.19$, so that one
rotation of the star corresponds to $5.26$ rotations of the inner
radius of the disk $P_*\approx 5.3 P_0$. Thus, to show the
variability pattern during few rotations of the star, we need
pretty long runs. Here, we show  sample results from several
relatively long runs with $t=15-16 P_0$. Figure 11 shows sample
light curves for the case of the cooler disk, and
$\Theta=15^\circ$ and $\Theta=45^o$. One can see that  the ``real"
light curves have similar features to the curves for fixed hot
spots. However, the variability pattern departs from the exact
variability patterns shown at Figures 7 and 8.  This is because
the hot spots continue to change their shape and position. Figure
12 shows variability curves for larger misalignment angles,
$\Theta=75^\circ$ and $\Theta=90^o$ (for warmer disk). One can see
that at these misalignment angles, the light curves are closer to
those of Figure 9, because the spots have almost fixed positions
on the stellar surface.

\begin{figure*}[t]
\epsscale{1.2} \plotone{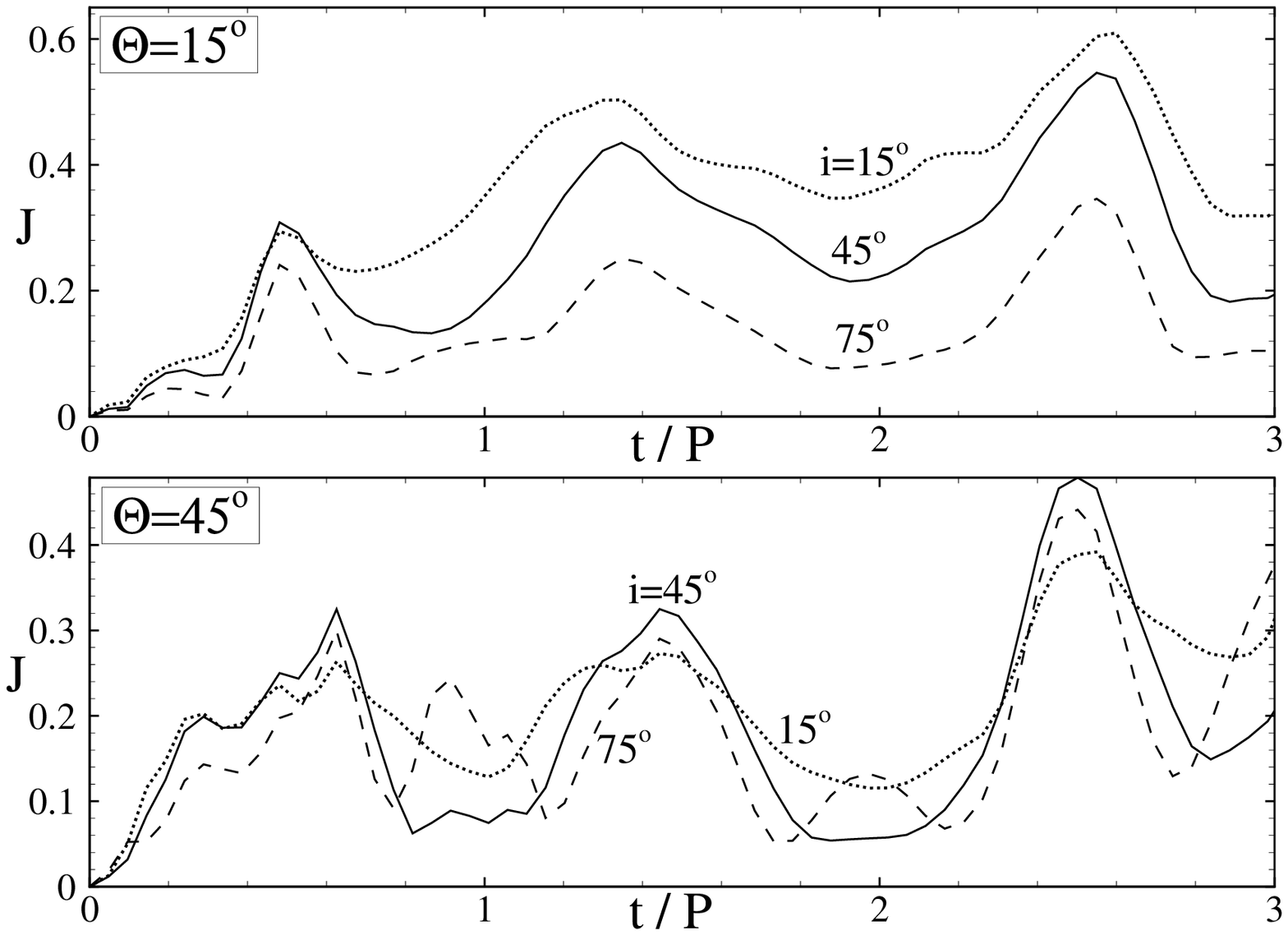} \caption{ Light curves calculated
for actual time-dependent hot spots for misalignment angles
$\Theta=15^\circ$ and $\Theta=45^\circ$, and for different
inclination angles $i$, for the case of the cooler disk,
$T_d=0.01$.} \label{Figure 11}
\end{figure*}
\begin{figure*}[t]
\epsscale{1.2} \plotone{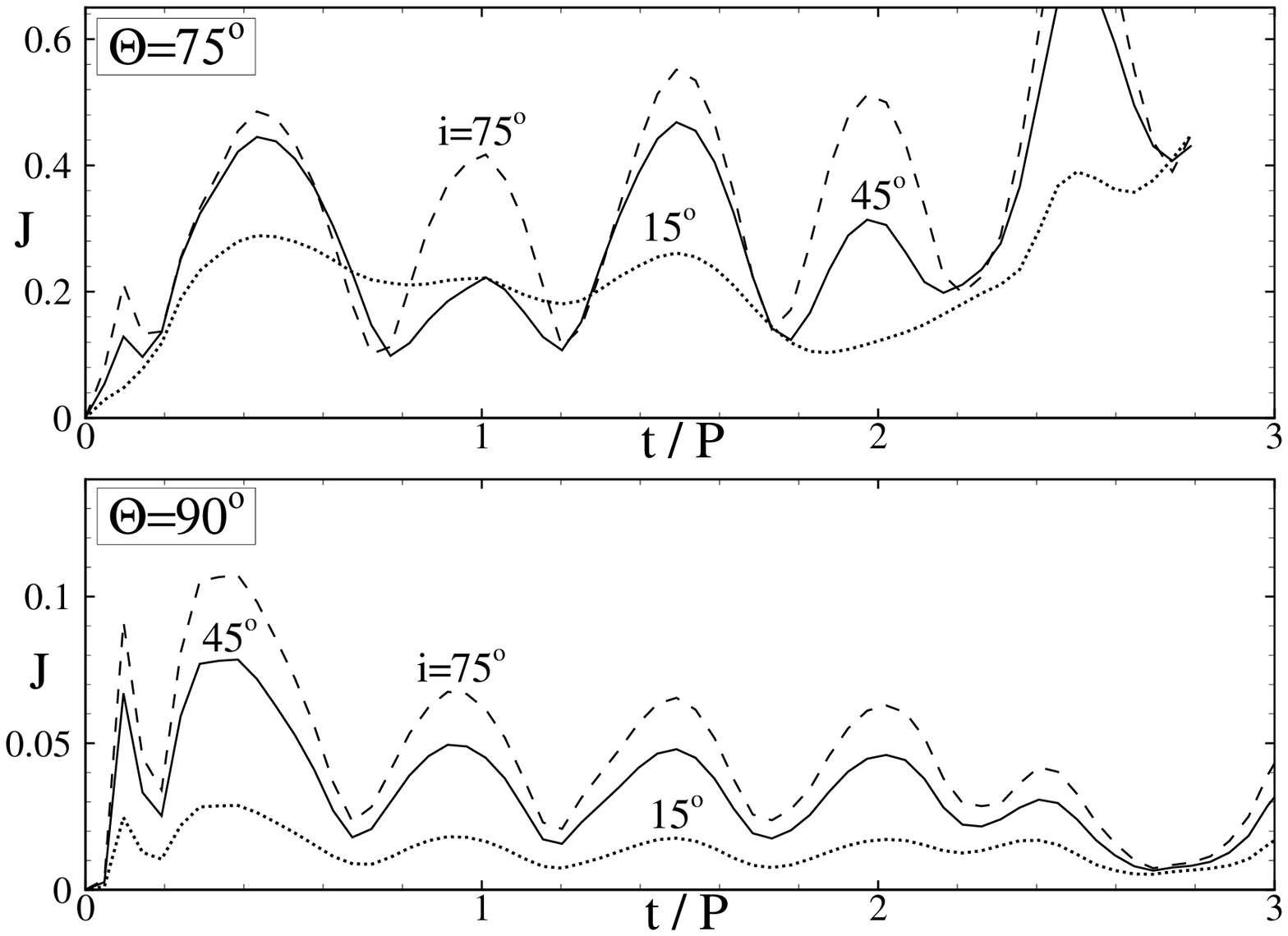} \caption{ Light curves calculated
for actual time-dependent hot spots for misalignment angles
$\Theta=75^\circ$ and $\Theta=90^\circ$, and for different
inclination angles $i$, for the case of the warmer disk,
$T_d=0.03$.} \label{Figure 12}
\end{figure*}

\section{Discussion}

In this paper we fixed parameters of the star and the flow. Below
we discuss dependence of the results on $\Omega_*$ and on the
accretion rate $\dot M$. In \S 5.2 we discuss the  limitations of
the model and future work.

\subsection{Dependence of Results on Parameters of the Star and
the Disk}

\noindent{\it $\Omega$--Dependence:} For the results given here we
have taken a relatively low angular velocity, $\Omega_*=0.19$.
This corresponds to CTTS with a period $P=9.4~{\rm days}$. We did
additional simulations for an intermediate angular velocity,
$\Omega_*=0.35$ ($P=3.3~{\rm days}$), and a rapidly rotating star,
$\Omega_*=1.0$ ($P=1.8~{\rm days}$). We observed that the hot spot
shapes are similar for slowly and for rapidly rotating stars. For
the intermediate angular velocity, the bow shape is less
prominent. The bow shape of the hot spots reflects the typical
shape of the funnel streams which have a thickness that is small compared
with their width. This is
 typical for relatively small inclination angles, $\Theta\lesssim
30^\circ$, where the stream should ``climb" an appreciable
distance above the equatorial plane before accreting to the
vicinity of the magnetic pole. At small inclination angles, the
bow shape is natural, because it may be considered as a part of
the narrow cylindrical ring which occurs for $\Theta=0$.

\begin{figure*}[t]
\epsscale{1.7} \plotone{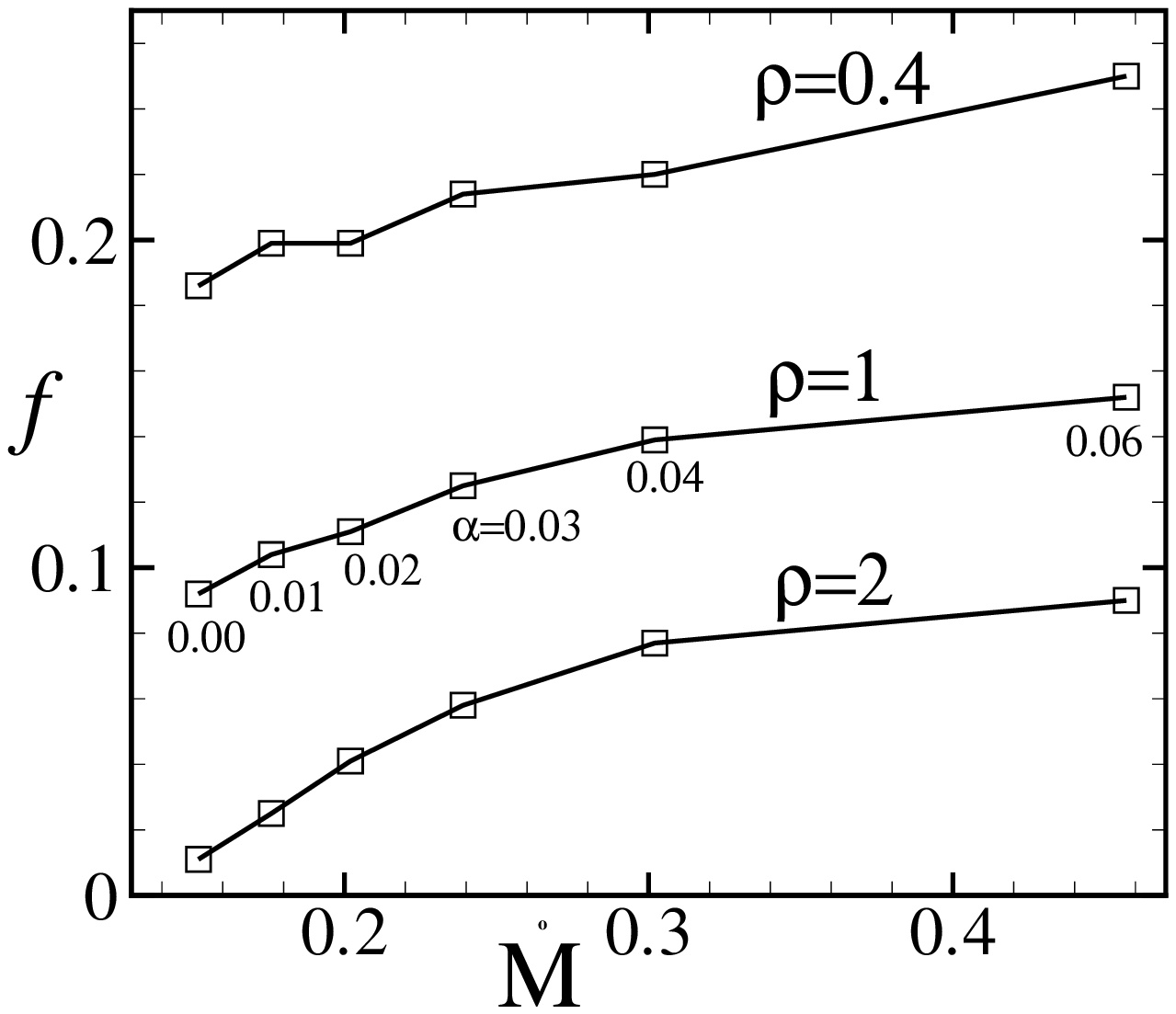} \caption {The fraction of the
star covered by the hot spots at the density levels: $\rho=0.4$,
$\rho=1$, and $\rho=2$ as a function of the accretion rate $\dot
M$. Squares show separate runs at different $\alpha$ parameters of
the disk viscosity.} \label{Figure 13}
\end{figure*}

\smallskip
\noindent{\it {$\dot M$--Dependence}}: We studied the dependence
of the size of the spots on the accretion rate $\dot M$. We
introduced an alpha-type viscosity analogous to one used in R02
 and performed a set of
runs at different values of accretion rate $\dot M$ (different
$\alpha$).
 Figure 13 shows that the fraction ${\it f}$ of the star
covered by the hot spots increases with the accretion rate for a range
of density levels, $\rho=0.4, 1, 2$. Thus, the size of the spots
increases as the accretion rate increases. This is in accord with
recent results of Ardila and Basri (2000), who analyzed the UV
variability of the CTTS BP Tau and have shown significant
correlation between the accretion rate and the filling factor of
the shocks. In R03 we noticed that at a larger accretion rate, the
streams become wider and cover a larger area of the magnetosphere.
This may possibly explain the observed variation of the shapes of
magnetospheric lines with the accretion rate (Muzerolle et al.
2001).

\subsection{Limitations of the Model and Future Work}

  Variability in different spectral lines in the accreting
magnetized stars may be associated with different regions: inner
regions of the disk, magnetospheric streams,  hot spots, or
outflows.
    In this paper
we analyze only the variability due to the rotation of the hot
spots,  while the variability associated with  {\it other regions}
will be investigated  in the future work.
  The light from the hot
spots may be also {\it obscured} by magnetospheric streams  or by
the warped inner regions of the disk if the star is approximately
edge-on (e.g., Bouvier, et al. 1999, 2003).
    This paper does not include the effect of obscuration.
    The warping of the inner regions of the disk
    (Aly 1980, Lipunov \& Shakura 1980, Lai 1999) was not
    observed in the simulations, but a special set of simulations
    will be done for investigation of this possible phenomenon.

 The magnetic field of the CTTS or millisecond pulsars may not be
{\it a pure dipole} field (Safier 1998; Smirnov et al. 2003).
   This can lead to more complicated geometry of the hot spots and more complicated
variability patterns.
    However, the detailed analysis of the
photometric and spectral variability of a number of CCTS  have
shown that most of the observed features can be explained by
models with a dipole magnetic field (e.g., Muzerolle, Hartmann, \&
Calvet 1998; Petrov, et al. 2001a,b; Alencar, Johns-Krull, \&
Basri 2001; Bouvier et al. 2003). This paper considers only the
case where the star's field is a dipole field. Non-dipolar field
geometries will be investigated separately.

We observed that the hot spots may rotate more rapidly or more
slowly than the star, which will lead to quasi-periodic
oscillations in the light curves and may explain QPOs observed in
CTTS (see Smith, Bonnell, \& Lewis 1995) and millisecond pulsars
(see, e.g., Chakrabarty et al. 2003). QPO variability in the
disk-magnetized star systems may also be associated with
oscillations of the inner radius of the disk (e.g., Goodson,
Winglee \& B\"ohm 1997). We plan to further investigate the
quasi-variability associated with such phenomena.

\section{Summary}

Disk accretion to a rotating star with a misaligned dipole
magnetic field has been studied further by three-dimensional MHD
simulations. This work focuses on the nature of the ``hot spots"
formed on the stellar surface due to the impact of two or more
funnel streams. We investigated the shape and intensity of the hot
spots for different misalignment angles $\Theta$ between the
star's rotation axis $\bf \Omega$ and its magnetic moment
$\rvecmu$. Further, we calculated the light curves due to rotation
of the hot spots for different angles $i$ between the observer's
line-of-sight and $\bf \Omega$. The main results are the
following:

\smallskip

\noindent{\bf 1.} For small inclination angles, $\Theta <
30^\circ$, the hot spots typically have a shape of a bow which is
bent around the magnetic pole. At large inclination angles,
$\Theta \gtrsim 60^\circ$, the shape becomes bar-like.  Often a
spot on a given hemisphere splits to form two spots, which
reflects the splitting of the funnel stream into two streams.
 The secondary stream is typically weaker than the
main stream so that one spot is much larger than the other.

\smallskip

\noindent{\bf 2.} The density, temperature, matter flux and other
parameters {\it increase towards the central regions of the
spots}, so that the spots are larger at lower temperature/density
and smaller at larger temperature/density. They cover about
$10-20\%$ of the area of the star at the density level typical for
the external regions of the funnel streams (see Figures 1 and 2).
The size of the hot spots increases with the accretion rate.

\smallskip

\noindent{\bf 3.} The spots have tendency to be located close to
the $\mu-\Omega$ plane. They tend to be located downstream of this
plane if a star rotates slowly (i.e., the inner region of the disk
and the foot-points of the stream rotate somewhat faster than the
star), or upstream, if a star rotates relatively fast. The spots
wander around their ``favorite" position. The amplitude of
wandering is smaller in case of the cooler disk.

\smallskip

\noindent{\bf 4.}  The calculated light curves reveal the
following features:

(a). The light curve has one peak per one period of rotation of
the star and the shape is approximately sinusoidal. This is
typical for small and medium misalignment angles, $\Theta \lesssim
45^\circ$, and inclination angles $i\lesssim 60^\circ$.~ (b). The
light curve has two peaks per period of rotation. This is typical
for all $\Theta$ if inclination angle is large, $i > 75^\circ$. At
very large misalignment angles, $\Theta\gtrsim 60^\circ-70^\circ$,
the double-peak curve is typical for wide range of inclination
angles, $i\gtrsim 30^\circ$.

\smallskip

\noindent{\bf 5.} The variation of the shape and location of the
spots will lead to departure from the exact variability and to
{\it quasi-variability}. At small misalignment angles, $\Theta <
30^\circ$, the streams (and hot spots) may rotate with velocity
different from that of the star (R03) thus leading to
quasi-periodic oscillations.

\smallskip

\acknowledgments This research was conducted using the resources
of the Cornell Theory Center, which receives funding from Cornell
University, New York State, Federal Agencies, foundations, and
corporate partners. This work was supported in part by NASA grants
NAG5-13220, NAG5-13060, and by NSF grant AST-0307817. AVK and GVU
were partially supported by grants INTAS CALL2000-491, RFBR
03-02-16548, by contract MIST \# 40.022.1.1.1106  and by Russian
program ``Astronomy''.
 The authors thank Drs. S.A. Lamzin and
P.P. Petrov for discussion of CTTS, Drs. D. Chakrabarty and D. Lai
for discussion of millisecond pulsars, to Dr. J. Stinchcombe for
editing the manuscript and to the referee for valuable
suggestions.

\end{document}